\documentstyle[12pt]{article}

\def\be{\begin{equation}}
\def\ee{\end{equation}}

\begin{document}

\begin{titlepage}
\setlength{\textwidth}{5.0in}
\setlength{\textheight}{7.5in}
\setlength{\parskip}{0.0in}
\setlength{\baselineskip}{18.2pt}
\setlength{\footskip}{0.5in}
\setlength{\footheight}{0in}

\renewcommand{\thefootnote}{\fnsymbol{footnote}}

%\hfill SOGANG-MP 5/13

\vspace{0.3cm}

\begin{center}
{\Large\bf Scientific understanding of the anisotropic universe in the warped products spacetime for aerospace power 
}
\end{center}
\vspace{2.3cm}

\begin{center}
Jaedong Choi\footnote{Electronic address: choijdong@gmail.com}
\par

\end{center}

\begin{center}

{Korea Air Force Academy, Cheongju 363-849, Korea}\par
\end{center}

\vskip 0.5cm
\begin{center}
{Jan 8, 2016}
\end{center}

\vfill

\begin{abstract}
We study the GMGHS spacetime to analyze anisotropic cosmology model which represents homogeneous but anisotropically expanding or contracting cosmology. In this paper we investigate the solution of GMGHS spacetime in form of doubly warped products possessing warping functions and find the Ricci curvature associated with three phases in the evolution of the universe. 
\end{abstract}

\vskip20pt

\end{titlepage}

\newpage
\thispagestyle{empty}
%%%%%%%%%%%%%%%%%%%%%%%%%%%%%%%%%%%%%%%%%%%%%%%%%%%%%%%%%%%%%%%%%%%%%%%
\section{ Introduction}
\setcounter{equation}{0}
\renewcommand{\theequation}{\arabic{section}.\arabic{equation}}
%%%%%%%%%%%%%%%%%%%%%%%%%%%%%%%%%%%%%%%%%%%%%%%%%%%%%%%%%%%%%%%%%%%%%%
 Since the cosmic microwave background was discovered, there have been many ideas and proposals to figure out how the universe has evolved. The standard big bang cosmological model based on the Friedmann-Robertson-Walker (FRW) spacetimes has led to the inflationary cosmology~\cite{guth} and nowadays to the M-theory cosmology with bouncing universes~\cite{seiberg}. \vskip5pt
The warped product manifold  was developed to point out that several of the well-known exact solutions to Einstein field equations are pseudo-Riemannian warped products~\cite{be}.  Furthermore, a general theory and Lorentzian multiply warped products were applied to discuss the Schwarzschild spacetime in the interior of the event horizon~\cite{choi00}.  From a physical point of view, these warped product spacetimes are interesting since they include classical examples of space-time such as the FRW manifold and the intermediate zone of Reissner-Nordstr\"{o}m (RN) manifold~\cite{rn,ksy}. The FRW cosmological model was studied to investigate non-smooth curvatures associated with multiple discontinuities involved in the evolution of the universe. Einstein postulated that the universe is homogeneous and isotropic at each moment of its evolution. The FRW metric is an exact solution of Einstein's field equations of general relativity. 
\vskip5pt
The GMGHS solution of the Einstein field equations represents the geometry exterior to a spherically symmetric static charged black hole and GMGHS metric in the doubly warped product spacetime has the same form with the Kantowski-Sachs solution~\cite{KS}. By turning antisymmetric tensor gauge fields off, the static charged black hole solution was found by Gibbons, Maeda~\cite{gm}, and by Garfinkle, Horowitz, Strominger~\cite{ghs}, independently. 
\vskip5pt
In this paper we will exploit $C^0$ scale function in the three phases of isotropic universe evolutio in the FRW metric. And we apply the GMGHS metric to analyze the anisotropic universe as a cosmological model, which can be treated as a doubly warped products spacetime having the Kantowski-Sachs solution which represents homogeneous but anisotropically expanding or contracting cosmology. Finally, we investigate the curvature in radiation-dominated(RD) universe in the anisotropic cosmology model by using the doubly warped product scheme.  We shall use geometrized units, i.e., G = c = 1, for notational
convenience.
\vskip20pt
%%%%%%%%%%%%%%%%%%%%%%%%%%%%%%%%%%%%%%%%%%%%%%%%%%%%%%%%%%%%%%%%%%%%%%%
\clearpage
\thispagestyle{empty}
\section{$C^{0}$ scale function in the three phases of isotropic universe evolution}
\setcounter{equation}{0}
\renewcommand{\theequation}{\arabic{section}.\arabic{equation}}

%%%%%%%%%%%%%%%%%%%%%%%%%%%%%%%%%%%%%%%%%%%%%%%%%%%%%%%%%%%%%%%%%%%%%%
The FRW metric describes a homogeneous, isotropic expanding cosmology. Isotropy means there are no special directions to the universe, and homogeneous means that there are no special places in the universe. In the spatially flat FRW cosmology with $k=0$, the early universe was radiation dominated, the adolescent universe was matter dominated, and the present universe is now entering into a dark energy dominated phase in the absence of vacuum energy. If the universe underwent inflation, there was a very early period when the stress-energy was dominated by vacuum energy. The Friedmann equation may be integrated to give the age of the universe in terms of present cosmological parameters. \vskip10pt
With the above astrophysical phenomenology in mind, consider the homogeneous but anisotropically expanding(contracting) cosmology in the two warped product spacetime. In physical cosmology, the radiation dominated era was the first of the three phases of the known universe, the other two being the matter dominated era and the dark energy dominated era. For a dark energy dominated universe the evolution of the scale factor in the FRW metric is obtained solving the Friedmann equations. 
\vskip10pt We have the scale factor $f$ as a function of time $t$ which scales as $f(t)\propto
t^{1/2}$ for a radiation dominated (RD)universe, and scales as $f(t)\propto t^{2/3}$ for a matter dominated (MD)universe, and scales as $f(t)\propto e^{Kt}$ for a dark energy dominated (DED)universe. Where the coefficient $K$ in the exponential, the Hubble constant, $K=\sqrt{8\pi G\rho_{full}/3}=\sqrt{\Lambda/3}$. This exponential dependence on time makes the spacetime geometry identical to the de Sitter Universe, and only holds for a positive sign of the cosmological constant, the sign that was observed to be realized in nature anyway. \vskip5pt

The current density of the observable universe is of the order of $9.44 \times 10^{-27}{kg m^{-3}}$ and the age of the universe is of the order of 13.8 billion years, or $4.358 \times 10^{17}s$. The Hubble parameter, $K$, is $\sim 70.88 km s^{-1}Mpc^{-1}$. (The Hubble time is 13.79 billion years.) The value of the cosmological constant, $\Lambda$, is $\sim 2 \times 10^{-35}s^{-2}$.
\clearpage
\thispagestyle{empty}
\vskip10pt
{\bf Definition  2.1}\ \ A $C^{0}$-Lorentzian metric on $M$ is a
nondegenerate (0,2) tensor of Lorentzian signature  such that\par
\hskip0.5cm$(i)$\ \ $g \in C^{0}$ on $S$\par 
\hskip0.45cm$(ii)$\ \ $g \in
C^{\infty}$ on $M\cap S^{c}$\par 
\hskip0.4cm$(iii)$\ \  For all $p \in
S$, and $U(p)$ partitioned by $S$, $g{\mid}_{U_{p}^{+}}$ and
$g{\mid}_{U_{p}^{-}}$ have smooth extensions to $U$. We call $S$ a
$C^{0}$-singular hypersurface of $(M, g)$.\vskip10pt
Consider $M_0 $ as a $C^{0}$-singular hypersurface of $(M, g)$. In
the spacetime, $f>0$ is smooth functions on
$M_0=(t_{0},\ t_{\infty})$ except at $t\not=t_{i}$ $(i=1,2)$, that
is $f\in C^{\infty}(S)$ (where $S=\{t_{i}\}\times_{f}H$) for
$t\not=t_{i}$ and $f\in C^{0}(S)$ at $t=t_{i}\in M_0$ to yield
\begin{equation}
f=\left(\begin{array}{l}
c_{0}t^{1/2},~~~~~~~~\mbox{$t<t_{1}$}\\
c_{1}t^{2/3},~~~~~~~~\mbox{$t_{1}\leq t\leq t_{2}$}\\
c_{2}e^{Kt},~~~~~~~~\mbox{$t>t_{2}$}
\end{array}
\right) \label{f012}
\end{equation}
with the boundary conditions
\begin{equation}
c_{0}t_{1}^{1/2}=c_{1}t_{1}^{2/3},~~~c_{1}t_{2}^{2/3}=c_{2}e^{Kt_{2}}.
\label{bc}
\end{equation}
Experimental values for $t_{1}$ and $t_{2}$ are given by
$t_{1}=4.7\times 10^{4}$ yr and $t_{2}=9.8$ Gyr.
Moreover $c_{1}$ and $c_{2}$ are given in terms of $c_{0}$,
$t_{1}$ and $t_{2}$ as follows
$$c_{1}=c_{0}t_{1}^{-1/6},~~~c_{2}=c_{0}t_{1}^{-1/6}t_{2}^{2/3}e^{-Kt_{2}}.$$
Thus we have 
\begin{equation}
f=\left(\begin{array}{l}
c_{0}t^{1/2},~~~~~~~~~~~~~~~~~~~~~\mbox{$t<t_{1}$}\\
c_{0}t_{1}^{-1/6}t^{2/3},~~~~~~~~~~~~~~~\mbox{$t_{1}\leq t\leq t_{2}$}\\
c_{0}t_{1}^{-1/6}t_{2}^{2/3}e^{K(t-t_{2})},~~~~~~\mbox{$t>t_{2}$}
\end{array}
\right) \label{f012}
\end{equation}
\vskip10pt
Here, we have the scale factor $f$ as a function of time $t$ for a radiation dominated (RD)universe, for a matter dominated (MD)universe, and for a dark energy dominated (DED)universe.

\vskip20pt
\clearpage
\thispagestyle{empty}

%%%%%%%%%%%%%%%%%%%%%%%%%%%%%%%%%%%%%%%%%%%%%%%%%%%%%%%%%%%%%%%%%%%%%%%

\section{The GMGHS metric of anisotropy spacetime ~\cite{choi04}}
\setcounter{equation}{0}
\renewcommand{\theequation}{\arabic{section}.\arabic{equation}}

%%%%%%%%%%%%%%%%%%%%%%%%%%%%%%%%%%%%%%%%%%%%%%%%%%%%%%%%%%%%%%%%%%%%%%
The GMGHS solution of the Einstein field equations represents the geometry exterior to a spherically symmetric static charged black hole. In the Schwarzschild coordinates, the line element for the GMGHS
metric in the exterior region $r>2m$ has the form as follows

\begin{equation}\label{rs}
ds^2=-\Bigl({1-\frac{2m}{r}}\Bigr)dt^2+\Bigl({1-\frac{2m}{r}}\Bigr)^{-1}dr^2
     +r^2\Bigl(1-{\frac{\alpha}{r}}\Bigr)(d\theta^2+\sin^2\theta d\phi^2),
\end{equation}
where
\begin{equation}
\label{r00w}e^{2\Phi}=1-\frac{\alpha}{r}, ~~F_{rt}=\frac{Q}{r^2}
\end{equation}
with $\alpha=Q^2/m$. Here, $\Phi$ is the dilaton field and $F_{rt}$ is the electric field strength. The parameters $m$ and $Q$ are mass and charge respectively. 
Note that the metric in the $t$-$r$ plane is identical to the Schwarzschild case. As in the Schwarzschild spacetime, the GMGHS has an event horizon at $r=2m$. We also note that the area of the sphere of the GMGHS
black hole, defined by $\int d\theta d\phi
\sqrt{g_{\theta\theta}g_{\phi\phi}}$, is smaller than the Schwarzschild spacetime by an amount depending on the charge. 
\vskip5pt 
In particular, the area of the sphere approaches zero as
$r\rightarrow\alpha$, leading to a surface singularity as
\begin{equation}
R=\frac{\alpha^2(r-2m)}{2r^3(r-\alpha)^2}.
\end{equation}
As far as the case of $\alpha\le 2m$, the singular surface remains inside the
event horizon so that the Penrose diagram is identical to the
Schwarzschild spacetime. Also the case of  $\alpha=2m$ implies
$Q^2=2m^2$ is the extremal limit at which the event horizon
and the surface singularity meet.
\vskip20pt
\clearpage
\thispagestyle{empty}
%%%%%%%%%%%%%%%%%%%%%%%%%%%%%%%%%%%%%%%%%%%%%%%%%%%%%%%%%%%%%%%%%%%%%%%

\section{GMGHS spacetimes in the form of doubly warped products~\cite{choi04}}
\setcounter{equation}{0}
\renewcommand{\theequation}{\arabic{section}.\arabic{equation}}

%%%%%%%%%%%%%%%%%%%%%%%%%%%%%%%%%%%%%%%%%%%%%%%%%%%%%%%%%%%%%%%%%%%%%%
In the paper~\cite{choi04}, considering the line element for the GMGHS metric, the interior region $r<2m$ can be described by \begin{equation}\label{inside}
ds^2=-\Bigl({\frac{2m}{r}}-1\Bigr)^{-1}dr^2+\Bigl({\frac{2m}{r}-1}\Bigr)dt^2
     +r^2\Bigl(1-{\frac{\alpha}{r}}\Bigr)(d\theta^2+\sin^2\theta d\phi^2),
\end{equation}
where $r$ and $t$ are now new temporal and spatial variables, respectively. A multiply warped product manifold, denoted by $M=(B\times F_1\times...\times F_{n}, g)$, consists of the Riemannian base manifold $(B, g_B)$ and fibers $(F_i,g_i)$ ($i=1,...,n$) associated with the Lorentzian metric~\cite{choi00}.
In particular, for the specific case of $(B=R,~g_B=-d\mu^{2})$,
the GMGHS metric (\ref{inside}) can be rewritten as two 
warped products $(a, b)\times_{f_1}R\times_{f_2} S^2$ by making
use of a lapse function
\begin{equation}
N^{2}=\frac{r_{H}-r}{r}
\label{schlapse2}
\end{equation}

as well as warping functions given by $f_1$ and $f_2$ as follows
\begin{eqnarray}\label{fs}
f_1(\mu)&=&\left(\frac{2m}{F^{-1}(\mu)}-1\right)^{1/2},\label{schf1f1}\\
f_2(\mu)&=&\left(F^{-1}(\mu)^2-\alpha F^{-1}(\mu)\right)^{1/2}.
\label{schf1f2}
\end{eqnarray}

The lapse function (\ref{schlapse2}) is well defined in the region $r<r_{H}(=2m)$ to rewrite it as two
warped products spacetime by defining a new coordinate $\mu$ as follows
\begin{equation}
\mu=\int_{0}^{r}\frac{dx~x^{1/2}}{(r_{H}-x)^{1/2}}=F(r).
\label{musch}
\end{equation}

Setting the integration constant zero as $r\rightarrow 0$, we have
\begin{equation}
\mu=2m\cos^{-1}\left(\frac{r_{H}-r}{r_{H}}\right)-[(r_{H}-r)r]^{1/2},
\label{schsolmu}
\end{equation}
which has boundary conditions as follows
\begin{eqnarray}
 \lim_{r\rightarrow r_{H}}F(r)=(2n-1)m\pi,~~~ \lim_{r\rightarrow 0}F(r)=0, \label{bdy2}
\end{eqnarray}
\clearpage
\thispagestyle{empty}
for positive integer $n$, and $dr/d\mu >0$ implies that
$F^{-1}(\mu)$ is well-defined function.  We can thus rewrite the
GMGHS metric (\ref{inside}) with the lapse function
(\ref{schlapse2})
\begin{eqnarray}\label{nsmetric2}
ds^{2}&=&-d\mu^2+\Bigl({\frac{2m}{F^{-1}(\mu)}}-1\Bigr)dr^2
          +\Bigl({F^{-1}(\mu)}^2-\alpha{F^{-1}(\mu)}\Bigr)d\Omega^2\nonumber\\
      &=&-d\mu^2+f_1(\mu)^2dr^2+f_2(\mu)^2d\Omega^2
\end{eqnarray}
by using the warping functions (\ref{schf1f2}). This GMGHS metric in the doubly warped product spacetime has the same form with the Kantowski-Sachs solution~\cite{KS} which represents homogeneous but anisotropically expanding or contracting cosmology. \vskip 5pt

In the case of the interior region $r<2m$, the GMGHS metric has
been rewritten as doubly warped product spacetime having two warping functions in terms of $f_1$ and $f_2$.
which have the same form with the Ricci curvature of the multiply
warped interior Schwarzschild metric~\cite{choi00}. The only
difference from the Schwarzschild is the $\alpha$ term in the
warping function $f_2(\mu)$ in Eq.~{(\ref{schf1f2})}. \vskip5pt

 Moreover, we can write
down the Ricci curvature on the two warped product of the GMGHS spacetimes as
\begin{eqnarray}
R_{tt}&=&-\frac{f_1''}{f_1}-\frac{2f_2''}{f_2},\nonumber\\
R_{rr}&=&f_1f_1''+\frac{2f_1f_1''}{f_2},\nonumber\\
R_{\theta\theta}&=&\frac{f_1'f_2f_2'}{f_1}+f_2'^2+f_2f_2''+1,\nonumber\\
R_{\phi\phi}&=&\left(\frac{f_1'f_2f_2'}{f_1}+f_2'^2+f_2f_2''+1\right)\sin^2\theta,\nonumber\\
R_{mn}&=&0,~{\rm for}~m\neq n,
\end{eqnarray}
\vskip10pt 
Anisotropy is the property of being directionally dependent, as opposed to isotropy, which implies identical properties in all directions. Now consider the curvature of the GMGHS anisotropy spacetimes.

\vskip20pt
\clearpage
\thispagestyle{empty}
\section{ Curvature of the GMGHS anisotropy spacetimes}
\setcounter{equation}{0}
\renewcommand{\theequation}{\arabic{section}.\arabic{equation}}
%%%%%%%%%%%%%%%%%%%%%%%%%%%%%%%%%%%%%%%%%%%%%%%%%%%%%%%%%%%%%%%%%%%%%%
In order to analyze the anisotropic universe as a cosmological model, which can be treated as a doubly warped products manifold possessing warping functions having the Kantowski-Sachs solution that represents homogeneous but anisotropically expanding(contracting) cosmology. Choose $\alpha=0$ in (\ref{inside}) to get the anisotropy spacetime. Then for a radiation-dominated (RD) universe era {~$t<t_{1}$}, put $$f_1(t)=c_{0}t^{1/2}$$  
we have 
\begin{equation}\label{rdf2}f_2(t)=F^{-1}(t)={\frac{2m}{c_0^2t+1}}\end{equation}
thus we have the metric of radiation-dominated (RD) universe era
\begin{equation}\label{rd1}ds^2=-dt^2+c_{0}^2tdr^2+\left({\frac{2m}{c_0^2t+1}}\right)^2d\Omega^2\end{equation}

\vskip10pt
For a matter-dominated (MD) universe era {~$t_{1}\leq t\leq t_{2}$}, put $$f_1(t)=c_{0}t_{1}^{-1/6}t^{2/3}=c_3t^{2/3}$$  
we have 
\begin{equation}\label{mdf2}f_2(t)=F^{-1}(t)={\frac{2m}{c_{3}^2t^{4/3}+1}}\end{equation}
thus we have the metric of matter-dominated (MD) universe era
\begin{equation}\label{mdf1}ds^2=-dt^2+c_{3}^2t^{4/3}dr^2+\left({\frac{2m}{c_{3}^2t^{4/3}+1}}\right)^2d\Omega^2\end{equation}
For a dark-energy-dominated (DED) universe era {~$t>t_{2}$}, put $$f_1(t)=c_{0}t_{1}^{-1/6}t_{2}^{2/3}e^{K(t-t_{2})}=c_4e^{Kt}$$  
we have 
\begin{equation}\label{ldf2}f_2(t)=F^{-1}(t)=\frac{2m}{c_{4}^2e^{2Kt}+1}\end{equation}
\clearpage
\thispagestyle{empty}

thus we have 
\begin{equation}\label{ld1}ds^2=-dt^2+c_{4}^2e^{2Kt}dr^2+\left(\frac{2m}{c_{4}^2e^{2Kt}+1}\right)^2d\Omega^2\end{equation}
\vskip10pt
Now we can write the curvature of a static spherically symmetric GMGHS spacetime in the radiation-dominated (RD) universe era as 

\begin{eqnarray}
&R_{tt}&=\frac{15c_0^4t^2-c_0^2t-1)}{t^2(4c_0^2 t+1)^2},\nonumber\\
&R_{rr}&=\frac{c_0^2(5c_0^2t+1)}{4t(c_0^2 t+1)},\nonumber\\
&R_{\theta\theta}&=-\frac{c_0^8t^5+4c_0^6t^4+10c_0^4m^2t+6c_0^4t^3-2c_0^2m^2+4c_0^2t^2+t}{(c_0^2t+1)^4},\nonumber\\
&R_{\phi\phi}&=-\frac{c_0^8t^5+4c_0^6t^4+10c_0^4m^2t+6c_0^4t^3_0-2c_0^2m^2+4c_0^2t^2+t}{(c_0^2t+1)^4}\sin^2\theta,\nonumber\\
&R_{mn}&=0,~{\rm for}~m\neq n,
\end{eqnarray}
\vskip20pt
and we also write the curvature of a static spherically symmetric GMGHS spacetime in the matter-dominated (MD) universe era as 
\begin{eqnarray}
&R_{tt}&=\frac{54c_3^4t^{10/3}-12c_3^2t^2-2t^{2/3}}{9t^{8/3}(c_3^2 t^{4/3}+1)^2},\nonumber\\
&R_{rr}&=\frac{18c_3^4t^{4/3}+2c_3^2}{9t^{2/3}(c_3^2 t^{4/3}+1)},\nonumber\\
&R_{\theta\theta}&=-\frac{m^2(48c_3^4t^{4/3}-16c_3^2)+3c_3t^6+12c_3^6t^{14/3}+18c_3^4t^{10/3}+12c_3^2t^2+3t^{2/3}}
{3t^{2/3}(c_3^2 t^{4/3}+1)^4},\nonumber\\
&R_{\phi\phi}&=-\frac{m^2(48c_3^4t^{4/3}-16c_3^2)+3c_3t^6+12c_3^6t^{14/3}+18c_3^4t^{10/3}+12c_3^2t^2+3t^{2/3}}
{3t^{2/3}(c_3^2 t^{4/3}+1)^4}\sin^2\theta,\nonumber\\
&R_{mn}&=0,~{\rm for}~m\neq n,
\end{eqnarray}
\vskip20pt
\clearpage
\thispagestyle{empty}
Finally  we also write the curvature of a static spherically symmetric GMGHS spacetime in the dark energy -dominated (DED) universe era as 

\begin{eqnarray}&R_{tt}&=\frac{k^2(9c_4^4(e^{2kt})^2 -6c_4^2e^{2kt}+1)}{c_4^2 e^{2kt}+1},\nonumber\\
&R_{rr}&=-\frac{c_4^2k^2e^{2kt}(3c_4^2 e^{2kt}-1)}{c_4^2 e^{2kt}+1},\nonumber\\
&R_{\theta\theta}&=-\frac{24m^2k^2(c_4^4e^{4kt}-c_4^2e^{2kt})+c_4^8 e^{8kt}+4c_4^6e^{6kt} +6c_4^4
e^{4kt}+4c_4^2e^{2kt}+1}{(c_4^2 e^{2kt}+1)^4},\nonumber\\
&R_{\phi\phi}&=-\frac{24m^2k^2(c_4^4e^{4kt}-c_4^2e^{2kt})+c_4^8 e^{8kt}+4c_4^6e^{6kt} +6c_4^4
e^{4kt}+4c_4^2e^{2kt}+1}{(c_4^2 e^{2kt}+1)^4}\sin^2\theta,\nonumber\\
&R_{mn}&=0,~{\rm for}~m\neq n,
\end{eqnarray}
\vskip20pt

%%%%%%%%%%%%%%%%%%%%%%%%%%%%%%%%%%%%%%%%%%%%%%%%%%%%%%%%%%%%%%%%%%%%%%%
\section{Conclusion}
\setcounter{equation}{0}
\renewcommand{\theequation}{\arabic{section}.\arabic{equation}}
%%%%%%%%%%%%%%%%%%%%%%%%%%%%%%%%%%%%%%%%%%%%%%%%%%%%%%%%%%%%%%%%%%%%%%

We have investigated a multiply warped product manifold associated with anisotropy spacetimes having the multiple discontinuities involved in the evolution of the universe and evaluated the Ricci curvature components. 
\vskip3pt 
Differently from the uncharged Schwarzschild metric where both the Ricci and Einstein curvatures vanish inside the horizon, the Ricci curvature components inside the GMGHS horizons are nonvanishing even though the Einstein scalar curvature vanishes in the interior of the charged for all step of evolution of the universe. 
\vskip3pt 
In fact, Ricci curvature is the part of the curvature of space-time that determines the degree to which matter will tend to converge or diverge. This describes not only homogeneous but also anisotropy at any point, and also demonstrates that the universe is expanding(contracting) at different rates in different directions. 

\section*{Acknowledgement}

J. Choi would like to acknowledge financial support from the Korea Air Force Academy Grant (KAFA 15-04).

\vskip10pt
\clearpage
\thispagestyle{empty}


\begin{thebibliography}{99}


\bibitem{be} J. K. Beem, P. E. Ehrlich and K. Easley, {\it Global Lorentzian Geometry} (Marcel Dekker Pure and Applied Mathematics, New York, 1996).
\bibitem{choi00} J. Choi, {\it  Multiply warped products with nonsmooth metrics}, J. Math. Phys. 41, 8163-8169 (2000).
\bibitem{choi01}  S.T. Hong, J. Choi and Y.J. Park, {\it (2+1) dimensional black holes in warped products}, Gen. Rel. Grav. 35 (2003).
\bibitem{choi02} J. Choi and S.T. Hong, {\it Warped product approach to universe with non-smooth scale factor}, J. Math. Phys. { 45}, 642 (2004).
\bibitem{choi03} S.T. Hong, J. Choi and Y.J. Park, {\it Local free-fall temperature of Gibbons-Maeda-Garfinkle-Horowitz-Strominger black holes} Nonlinear Analysis {\bf 63}, e493 (2005).
\bibitem{choi04} J. Choi {\it The geodesic motion near hypersufaces in the warped products spacetime} Nonlinear Analysis {\bf 28}, 32 (2013).
\bibitem{ksy} J. Demers, R. Lafrance, and R.C. Meyers, {\it Black hole entropy without brick walls}, Phys. Rev. D52, 2245-2253 (1995).
\bibitem{ksy1} A. Ghosh and P. Mitra, {\it Entropy for extremal Reissner-Nordstrom black holes}, Phys. Lett. B357, 295-299 (1995).
\bibitem{ksy2} S.P. Kim, S.K. Kim, K.S. Soh and J.H. Yee, {\it Remarks on renormalization of black hole entropy}, Int. J. Mod. Phys. A12, 5223-5234 (1997).
\bibitem{ksy3} G. Cognola and P. Lecca, {\it Electromagnetic fields in Schwarzschild and Reissner-Nordstrom geometry}, Phys. Rev. D57, 1108-1111 (1998).
\bibitem{ghs} D. Garfinkle, G.T. Horowitz and A. Strominger, {\it Hawking radiation from Garfinkle-Horowitz-Strominger and nonextremal D1-D5 black holes via covariant anomalies}, Phys. Rev. D { 43}, 3140 (1991).
\bibitem{gm} G.W. Gibbons and K. Maeda, Nucl. {\it Energy Associated with the Gibbons-Maeda Dilaton Spacetime}Phys. B { 298}, 741 (1988).
\clearpage
\thispagestyle{empty}
\bibitem{guth} A.H. Guth, {\it The inflationary universe: a possible solution to the
horizon and flatness problems}, Phys. Rev. D23, 347-356 (1981).
\bibitem{KS} R. Kantowski and R.K. Sachs, {\it Some spatially inhomogeneous dust models}, J. Math. Phys. {7}, 443 (1966).
\bibitem{seiberg} J. Khoury, B.A. Ovrut, N. Seiberg, P.J. Steinhardt and N. Turok,
{\it From big crunch to big bang}, Phys. Rev. D65, 086007 (2002).
\bibitem{rn} H. Reissner, {\it \"Uber die eigengravitation des elektrischen felds nach der
Einsteinshen theorie}, Ann. Phys. 50, 106-120 (1916).
\bibitem{rn1} G. Nordstr\"om, {\it On the energy of the gravitational 
field in Einstein's theory}, Proc. Kon. Ned. Akda. Wet. 20, 1238-1245
(1918). 
\bibitem{Wald} R. M. Wald, {\it General Relativity} (University of Chicago Press, Chicago, 1984).
\end{thebibliography}
\end{document}